\documentclass{iopart}

\eqnobysec

\begin{document}

\jl{1}

\title{An alternative field theory for the
Kosterlitz-Thouless transition}

\author{Georg Foltin}
\address{Institut f\"ur Theor\-etische
Physik, Hein\-rich\---Heine--Uni\-ver\-sit\"at D\"us\-sel\-dorf,
Uni\-ver\-sit\"ats\-strasse
1, D--40225 D\"ussel\-dorf, Germany}

\begin{abstract}
We  extend a Gaussian model for the internal electrical potential of a two-dimensional Coulomb gas by
a non-Gaussian measure term, which singles out the physically relevant
configurations of the potential.
The resulting Hamiltonian, expressed as a functional of the internal potential, has a surprising
large-scale limit: The additional term simply counts the number of maxima and minima of the
potential. The model allows for a transparent derivation of the divergence of the
correlation length upon lowering the temperature down to the Kosterlitz-Thouless transition
point.
\end{abstract}

\pacs{6460, 6130J, 6470}

\section{Introduction}

The two-dimensional Coulomb gas has found an abundant number of applications, ranging from
the melting of two-dimensional crystals \cite{Hal78,You79},
vortices in superfluid films \cite{Bis78} and thin superconductors
\cite{Bea79} and
arrays of Josephson contacts \cite{Abr82} to
topological defects of thin liquid-crystal films \cite{Gen93}.
Common to these systems is the existence of two types of topological,
point-like defects (positive and
negative `charges'), which interact according to the two-dimensional Coulomb
law, i.e. the potential energy grows logarithmically with distance.
At low temperatures positive and negative defects are bound together
and form neutral dipoles, which start to unbind at a certain critical
temperature -- the Kosterlitz-Thouless transition point \cite{Kos72,Min87}.
In fact, an energy  of the order of $\Delta E\sim\log A$ ($A$ is the system size) is needed to
break up two bound charges. On the other hand, the unbinding process implies a gain of entropy,
which is also of the order of $\log A$. Above the critical temperature entropy wins and a
plasma-like gas of free charges is formed.
The high-temperature phase is characterized by the existence of a screening length:
The Coulomb interaction is shielded beyond the so called Debye-H\"uckel length due to the
formation of a cloud of opposite charges
around a (test-) charge.
This  phase is well described by a Gaussian model for the continuous
charge density $\rho$ (Debye-H\"uckel model) with the Hamiltonian
\begin{equation}
\label{DH}
\mathcal{H}/T=\frac{K_A}{2}\int \rmd^2x\,\rmd^2x'\,
\rho(\bi{x})\,G(\bi{x}-\bi{x}')\,\rho(
\bi{x}')+\int \rmd^2x\,\frac{\rho^2}{2z},
\end{equation}
where $G_{\bi{r}}\sim-(2\pi)^{-1}
\log(r/\ell)+$const
($\ell$ is a microscopic lengthscale) is the Green function of the two-dimensional Laplacian
$(-\nabla^2 G(r) = \delta(\bi{r}))$, $K_A$ is the charge-charge coupling
strength and the mass $z$ measures the distance to the transition
temperature $z\sim T-T_c$ and controls the density of charge pairs.
The charge--charge correlation function reads (in Fourier space)
\begin{equation}
C(p)\propto\left(1-\frac{K_Az}{p^2+K_Az}\right)
\end{equation}
where the screening length  (Debye-H\"uckel length $\ell_{DH}$) is given by
$\ell_{DH}=(K_Az)^{-1/2}$.
The Gaussian functional (\ref{DH}) applies to temperatures well above the transition
temperature - it does not predict the essential singularity of the correlation length
$\xi\sim\exp(b/\sqrt{T-T_c})$ closer to the transition point.

The aim of this paper is to propose a novel model Hamiltonian for the
high temperature phase of the two-dimensional Coulomb gas, describing
the fluctuations of the internal electrical potential. Contrary to the
sine--Gordon theory, which has an unphysical auxiliary variable as its
fundamental field, our model is based on a physical quantity (the potential)
and, therefore, leads to a more intuitive understanding of the
Kosterlitz--Thouless transition.

The resulting continuum model displays the correct divergence of the correlation length in the vicinity of the critical point, as will be
demonstrated within a variational calculation and,
in addition, within renormalized perturbation theory.

\section{The model}
We start with the (Gaussian) Debye-H\"uckel model (\ref{DH}) and introduce the internal electrical
potential $\phi$ via $-\nabla^2\phi=\rho$. The partition function for the model reads

\begin{equation}
\mathcal{Z}=\int D[\phi]\,\mathcal{W}[\phi]\exp\left(
-\frac{K_A}{2}\int \rmd^2x(\nabla\phi)^2-\frac{\mu}{2}\int \rmd^2x(\nabla^2\phi)^2
\right)
\end{equation}
with the (trivial) measure  $\mathcal{W}[\phi]\equiv 1$.
Can we do better and find an appropriate measure $\mathcal{W}[\phi]$ for the
internal potential, which singles out the relevant
configurations
and, therefore, go beyond the Gaussian case?
Here is a suitable candidate:
\begin{equation}
\label{measure}
\mathcal{W}[\phi]=\exp\left(-\frac{u}{4}\int \rmd^2x
(\nabla^2\phi)^2\,\delta_\lambda(\nabla\phi)\right)\equiv\exp\left(-\mathcal{H}_1/T\right)
\end{equation}
with the Gaussian (`delta function') $\delta_\lambda(\bi{E})=(2\pi\lambda)^{-1}\exp(-E^2/(2\lambda))$.

First, we investigate the \textit{mathematics} of this ansatz.
For $\lambda\rightarrow\infty$ the measure $\mathcal{W}[\phi]$ is Gaussian and simply
renormalizes the mass $z$. For small $\lambda$, on the other hand,
the function $\delta_\lambda(\nabla\phi)$ picks up zeros of $\nabla\phi$, i.e. local
maxima, minima and saddle points of the potential $\phi$.
We expand $\phi$ in the vicinity
of an extremal point $\bi{x}_0$ (where $\nabla\phi(\bi{x}_0)=0$)
 up to second order in
$\bi{\xi}=\bi{x}-\bi{x}_0$:
\[\phi(\bi{x}_0+\bi{\xi})\approx\phi(\bi{x}_0)
+(1/2)\alpha_{ij}\xi_i\xi_j\qquad i,j=1,2\] (summation over double indices is
assumed) and find for the
contribution of this extremum (for $\lambda\rightarrow 0$)
\begin{eqnarray}
\mathcal{H}_1/T&=&
\frac{u}{4}\int \rmd^2x(\nabla^2\phi)^2\,\delta_\lambda(\nabla\phi)\nonumber\\
&\approx&
\frac{u}{4}\int \rmd^2\xi\,(\alpha_{ii})^2\frac{1}{2\pi\lambda}\exp\left(
-\frac{\alpha_{ij}\alpha_{ik}\xi_j\xi_k}{2\lambda}\right)\nonumber\\
&=&
u\left(\frac{\eta_1+\eta_2}{2}
\right)^2/\left|\eta_1\eta_2\right|
\end{eqnarray}
where $\eta_{1,2}$ are the eigenvalues of the second derivative
$\partial_i\partial_j\phi=\alpha_{ij}$ right at the extremal point.
 Maxima/minima yield a contribution $\mathcal{H}_1/T\ge u$, since $\eta_1\cdot
\eta_2>0$. For a symmetric maximum/minimum $\eta_1=\eta_2$ and
consequently $\mathcal{H}_1/T= u$; on the other hand,
a `symmetric' saddle point yields zero, since $\eta_1=-\eta_2$.
Therefore, the nonlinear term acts as a kind of chemical potential for the number of
extremal points \cite{Hal81,Wei82,Fol00}%
\footnote{The number of saddles is equal to the number of maxima and minima for
a flat geometry.}:
\begin{equation}
\mathcal{H}_1/T\ge u\times\#(\mbox{local maxima and minima of }\phi).
\end{equation}
To summarize this point, the nontrivial measure $\mathcal{W}[\phi]$ favours
potentials with only few maxima and minima, which are in fact the physically
relevant ones. Consider a typical charge configuration of a Coulomb gas system
right above the transition temperature. Apart from a large number of dipols,
which give rise to an effective dielectric constant, we find \textit{few} free
charges $q_i=\pm 1$ at positions $\bi{r}_i$. The corresponding potential
$\phi(\bi{r})=\sum_iq_iG(\bi{r}-\bi{r}_i)$ has (few) maxima and minima at the
locations of the charges $\bi{r}_i$ (if we use a short-distance regularized
Green function $G$, otherwise the potential $\phi$ would diverge at the
locations $\bi{r}_i$) and symmetric saddle points elsewhere, since
$\Delta\phi(\bi{r})=0$ for $\bi{r}\neq\bi{r}_i$.

We rescale the potential $\sqrt{\mu}\phi\rightarrow\phi$ and the constant $\mu\lambda\rightarrow\lambda$, define the mass $\tau\equiv K_A/\mu$,
and obtain as the final model
\begin{equation}
\label{model}
\mathcal{H}/T=\int \rmd^2x\left(\frac{1}{2}(\nabla^2\phi)^2+\frac{\tau}{2}(\nabla\phi)^2
+\frac{u}{4}(\nabla^2\phi)^2\,\delta_\lambda(\nabla\phi)\right).
\end{equation}
The coupling constants $u,\lambda$ are dimensionless, whereas  $\tau$ is a relevant
coupling  with dimension $\tau\sim\mbox{length}^{-2}$--it is a measure
for the deviation from the critical point $\tau\sim T-T_c$.

\section{Variational approach}
\label{vari}

We calculate an upper bound of the free energy $\cal F$ of our model (\ref{model})
with the help of
\begin{equation}
\mathcal{F}=-\log\int D[\phi]\exp(-\mathcal{H}/T)\le \mathcal{F}_v+
\left<\mathcal{H}-\mathcal{H}_v\right>_v
 \end{equation}
and the ansatz
\begin{equation}
\label{vario}
\mathcal{H}_v/T=\frac{A}{2}\int \rmd^2x \left((\nabla^2\phi)^2
+\omega(\nabla\phi)^2\right)
\end{equation}
where $A$ and $\omega$ are variational parameters, $\mathcal{F}_v$ is the free energy with respect
to $\mathcal{H}_v$ and $\langle\ldots\rangle_v$ denotes the corresponding average.
In fact, the optimal Gaussian fit has the form given above---a more general
ansatz is not necessary and would reduce to an expression such as (\ref{vario}).
To evaluate $\left<(\nabla^2\phi(\bi{r}))^2\,\delta_\lambda(\nabla\phi(\bi{r}))\right>_v$ we note that $\nabla\phi(\bi{r})$ and $\nabla^2\phi(\bi{r})$ are uncorrelated
Gaussian variables and, therefore (we drop the argument $\bi{r}$ from now on),
\begin{eqnarray}
\fl \left<(\nabla^2\phi)^2\,\delta_\lambda(\nabla\phi)\right>_v
=\left<(\nabla^2\phi)^2\right>_v
\int \rmd^2E\,\frac{\exp\left(-\bi{E}^2/\left<(\nabla\phi)^2\right>_v
-\bi{E}^2/(2\lambda)
\right)}{\pi\left<(\nabla\phi)^2\right>_v\,2\pi\lambda}
\nonumber\\
\lo{=}
\frac{\left<(\nabla^2\phi)^2\right>_v}{\pi\left<(\nabla\phi)^2\right>_v+2\pi
\lambda}.
\end{eqnarray} The expectation value $\left<(\nabla\phi)^2\right>_v$
diverges for $\omega\rightarrow 0$ and, therefore, the width $\lambda$ of the Gaussian
$\delta_\lambda$ becomes irrelevant, justifiying the introduction of a sharp delta-function
($\lambda\rightarrow 0$) in the model (\ref{model}).
Up to the constant $\left<\mathcal{H}_v\right>_v$, we obtain the upper bound of the free energy per area $f$

\begin{eqnarray}
\fl 8\pi f\le\int_0^{\Lambda^2}\rmd s\,\log(A(s+\omega))
+\int_0^{\Lambda^2}\rmd s \frac{s^2}{As(s+\omega)}+\tau\int_0^{\Lambda^2}\rmd s
\frac{s}{As(s+\omega)}\nonumber\\
+2u\int_0^{\Lambda^2}\rmd s\frac{s^2}{As(s+\omega)}\left(
\int_0^{\Lambda^2}\rmd s\frac{s}{As(s+\omega)}\right)^{-1}
\end{eqnarray}
where the first term represents $\mathcal{F}_v$, the second and third term are the expectation value of
the Gaussian
part of (\ref{model}) and the last term represents $\left<\mathcal{H}_1\right>_v$.
$\Lambda$ is the upper cutoff momentum and $s=p^2, \pi\rmd s=\pi p \rmd p =\rmd^2p$
using polar coordinates.
Next, we set $\Lambda=1$ (equivalently we can introduce dimensionless
couplings), eliminate $A$ (variation of the bound $f$ with respect to $A$
yields $A=1+(\tau-\omega)\log(1+1/\omega)$)
and arrive at
\begin{eqnarray}
\label{freeenergy}
\fl 8\pi f\le\log\left(1+(\tau-\omega)\log(1+1/\omega)\right)+\log(1+\omega)+
\omega\log(1+1/\omega)\nonumber\\
+2u\left(\frac{1}{\log(1+1/\omega)}-\omega\right).
\end{eqnarray}
It can be tested afterwards that $\omega$ and
$(\tau-\omega)\log(1+1/\omega)$ become small enough in the critical region
to approximate
$\log(1+(\ldots))\approx(\ldots)$ in the first two terms of (\ref{freeenergy}).
The expression is especially simple for $u=1/2$
\begin{equation}
8\pi f\le\tau\log(1+1/\omega)+\frac{1}{\log(1+1/\omega)}
\end{equation}
yielding the minimum $\log\left(1+1/\omega\right)=1/\sqrt{\tau}$ or
\begin{equation}
\omega\approx\exp\left(-\frac{1}{\sqrt{\tau}}\right)
\end{equation}
where we have used $1+\omega^{-1}\approx\omega^{-1}$ for $\tau\rightarrow 0$.
$\xi=\omega^{-1/2}$ is the correlation length of the best Gaussian fit, hence
an estimate for the correlation length as a function of $\tau$
\begin{equation}
\xi\sim\exp\left(\frac{1}{2\sqrt{\tau}}\right)
\end{equation}
which is the celebrated essential singularity of the correlation length
in the vicinity of the Kosterlitz--Thouless transition. For general $u$
we obtain $\omega=\exp(-\sqrt{2u/\tau})$.

\section{Renormalized perturbation theory}

We have also studied the model Hamiltonian (\ref{model}) within a simple (renormalized)
perturbation expansion. Using the formalism presented in
\cite{Hal81,Wei82,Fol00},
we have calculated the two-point vertex function $\Gamma_2(p)$,
which is the reciprocal Fourier transform of the
Green  function $\langle\phi(\bi{x})\phi(\bi{y})\rangle$, to
lowest order in the coupling constant $u$. Details of the derivation can be found in the appendix.
The vertex function displays a highly unconventional large-scale
behaviour if compared to usual critical field theory. The resulting effective 
couplings are in fact finite in the limit $\tau\rightarrow 0$ and fixed cutoff.
Since an instability shows up as a mere artefact of the expansion, we have to introduce
a (finite) shift of the mass $\tau\rightarrow\tau+u\,\Delta\tau$, where we
treat $u\,\Delta\tau \int \rmd^2x(\nabla\phi)^2/2$ along with the interaction
term as a perturbation.
For fixed spatial dimension $d=2$, fixed cutoff momentum $\Lambda=1$, a sharp delta function $\lambda
\rightarrow 0$
and up to $\Or (u^2)$, the two-point vertex function reads (see appendix)
\begin{equation}
\Gamma_2(p)=A_{\mbox{\tiny eff}}\left(p^4+\tau_{\mbox{\tiny eff}}\,p^2\right)
\end{equation}
with the effective couplings
\begin{eqnarray}
\tau_{\mbox{\tiny eff}}&=&\tau+u\,\Delta\tau-\frac{2u}{\left(\log(1+1/\tau)\right)^2}\\
A_{\mbox{\tiny eff}}&=&1+\frac{2u}{\log(1+1/\tau)}.
\end{eqnarray}
As a consequence of the $\delta(\nabla\phi)-$term and in contrast to ordinary field theories,
the $\log-$terms show up in the denominator and, consequently, the effective couplings remain
finite in the critical limit $\tau\rightarrow 0$. However, at a particular $\tau$, provided
$\Delta\tau=0$, the
effective mass becomes negative, signalling the breakdown of the naive perturbation theory.
Even worse, the correction to the mass, divided by $\tau$ itself, tends to infinity
in the limit $\tau\rightarrow 0$ . To absorb this divergence, we renormalize the mass by setting
\begin{equation}
\Delta\tau=\frac{2}{\left(\log(1+1/\tau)\right)^2}.
\end{equation}
Now, the effective mass reads $\tau_{\mbox{\tiny eff}}=\tau$ and the deviation from the critical
point is $\tau_0=\tau+u\,\Delta\tau\approx u\,\Delta\tau$ for small $\tau$. We arrive at
\begin{equation}
\tau_0=\frac{2u}{\left(\log(1+1/\tau_{\mbox{\tiny eff}})\right)^2}
\end{equation}
or $\tau_{\mbox{\tiny eff}}\sim\exp\left(-\sqrt{2u/\tau_0}\right)$
in agreement with the variational calculation.

\section{Conclusions}

We have proposed an alternative model for the
high-temperature phase of the two-dimensional Coulomb gas.
It describes the fluctuations of the internal electrical potential with the help of a measure term, which
favours potentials corresponding to configurations of only few charges.
The range of validity of
this model apparently extends down to the critical point---the model yields
the correct singular behaviour of the correlation length while approaching the
Kosterlitz--Thouless point from above, as shown within a simple
(renormalized) perturbation expansion. In addition,
a Gaussian variational approximation scheme turned out to be successful,
in contrast to the sine--Gordon theory \cite{Ami80},
where an analogous Gaussian variational ansatz
fails and yields a wrong singularity for the correlation length \cite{Sai78}.

Several questions could not be addressed,
for example how to extract the singular
behaviour of the free energy or the universal critical superfluid density.
In our opinion, the model deserves future investigations---thanks to its
fascinating and unconventional properties.

\ack

It is a pleasure to acknowledge discussions with R Blossey,
R Bausch, H K Janssen, B Schmittmann,
R K P Zia and H A Pinnow.
This work has been supported by the Deutsche Forschungsgemeinschaft
under SFB 237.

\appendix

\section{Calculation of the two-point vertex function}

For fixed spatial dimension $d=2$, $\lambda\rightarrow 0$ and up to $\Or (u^2)$,
the two-point correlation function reads
\begin{eqnarray}
\lefteqn{
\left<\phi(\bi{x})\phi(\bi{y})\right>=\left<\phi(\bi{x})\phi(\bi{y})\right>_0
}\\
&&\mbox{}+\frac{u}{4}\int \rmd^2z\left(\left<\phi(\bi{x})\phi(\bi{y})\right>_0
\left<2\,\Delta\tau\left(\nabla\phi(\bi{z})\right)^2+\left(\nabla^2\phi(\bi{z})\right)^2\,\delta\left(
\nabla\phi(\bi{z})\right)
\right>_0\right.\nonumber\\
&&\left.\mbox{}-\left<\phi(\bi{x})\phi(\bi{y})\left(
2\,\Delta\tau\left(\nabla\phi(\bi{z})\right)^2+\left(\nabla^2\phi(\bi{z})\right)^2\,
\delta\left(\nabla\phi(\bi{z})\right)\right)
\right>_0\right)+\Or (u^2)\nonumber
\end{eqnarray}
where $\langle\ldots\rangle_0$ denotes the (Gaussian) average with respect to
the quadratic part
$\mathcal{H}_0=(1/2)\int \rmd^2x((\nabla^2\phi)^2+\tau(\nabla\phi)^2)$
of the Hamiltonian (\ref{model}).
As explained in the main text, we have introduced a mass shift
$\tau\rightarrow\tau+u\,\Delta\tau$ and treat $u\,\Delta\tau \int
\rmd^2x(\nabla\phi)^2/2$ as a counterterm (additional perturbation). The bare
Green function reads \begin{equation}
G_0(p)=\int \rmd^2x \left<\phi(\bi{x})\phi(0)\right>_0\exp\left(-\rmi\bi{p}\cdot\bi{x}\right)
=\frac{1}{p^2(p^2+\tau)}.
\end{equation}
The counterterm adds
\begin{equation}
G_c(p)=-u\,\Delta\tau\,G_0(p)^2p^2=-u\,\Delta\tau\,\frac{1}{p^2(p^2+\tau)^2}.
\end{equation}
More involved are the contributions from the measure.
We define the Gaussian variables
$\phi_1\equiv\phi(\bi{x}),
\phi_2\equiv\phi(\bi{y}), \rho\equiv\nabla^2\phi(\bi{z}), \bi{E}\equiv
\nabla\phi(\bi{z})$, the five-component vector $M\equiv(\phi_1,\phi_2,\rho,\bi{E})$,
and write down the distribution of $M$ with the help of a Gaussian transformation \cite{Fol00}
\begin{equation}
P(M)=\frac{1}{(2\pi)^5}\int \rmd^5\tilde{M}\exp\left(-\frac{1}{2}\sum_{i,j=1}^5C_{ij}\tilde{M}_i
\tilde{M}_j+\rmi\tilde{M}\cdot M\right)
\end{equation}
where $C_{ij}=\left<M_iM_j\right>_0$ denotes the correlation matrix.
Explicitly, we have (we drop the subscript $0$ from now on)
\begin{eqnarray}
\lefteqn{(*)\equiv\left<\phi(\bi{x})\phi(\bi{y})\left(
\nabla^2\phi(\bi{z})\right)^2\,
\delta\left(\nabla\phi(\bi{z})\right)\right>}\nonumber\\
&=&\frac{1}{(2\pi)^5}\int \rmd\tilde{\phi}_1\rmd\tilde{\phi}_2
\rmd\tilde{\rho}\,\rmd^2\tilde{\bi{E}}\,\rmd\phi_1\rmd\phi_2\rmd\rho\,\rmd^2\bi{E}\nonumber\\
&&\times\exp\left(-\frac{1}{2}\left(\left(\tilde{\phi}_1^2+\tilde{\phi}_2^2\right)\left<\phi^2\right>+
\tilde{\rho}^2\left<(\nabla^2\phi)^2\right>+\tilde{\bi{E}}^2\left<(\nabla\phi)^2
\right>/2\right.\right.\nonumber\\
&&\left.\mbox{}+2\tilde{\phi}_1\tilde{\phi}_2\left<\phi(\bi{x})\phi(\bi{y})\right>
+2\tilde{\rho}\tilde{\phi}_1\left<\phi(\bi{x})\nabla^2\phi(\bi{z})\right>
+2\tilde{\rho}\tilde{\phi}_2\left<\phi(\bi{y})\nabla^2\phi(\bi{z})\right>\right.\nonumber\\
&&\left.\mbox{}+2\tilde{\bi{E}}\tilde{\phi}_1\cdot\left<\phi(\bi{x})\nabla\phi(\bi{z})\right>
+2\tilde{\bi{E}}\tilde{\phi}_2\cdot\left<\phi(\bi{y})\nabla\phi(\bi{z})\right>
\right)\nonumber\\
&&\left.\mbox{}+\rmi\left(\tilde{\phi}_1\phi_1+\tilde{\phi}_2\phi_2+\tilde{\rho}\rho+\tilde{\bi{E}}\cdot
\bi{E}\right)\right)\phi_1\phi_2\rho^2\delta(\bi{E})
\end{eqnarray}
where the correlations
$\left<\nabla^2\phi(\bi{z})\nabla\phi(\bi{z})\right>$ vanish by symmetry.
Next, we perform the trivial $\bi{E}$ integration and the $\tilde{\bi{E}}$ integration and obtain
\begin{eqnarray}
\label{reduc}
(*)&=&\frac{1}{(2\pi)^3\,\pi\left<(\nabla\phi)^2\right>}\int\rmd\tilde{\phi}_1\rmd\tilde{\phi}_2
\rmd\tilde{\rho}\,\rmd\phi_1\rmd\phi_2\rmd\rho\nonumber\\
&&\times\exp\left(-\frac{1}{2}\left(\left(\tilde{\phi}_1^2+\tilde{\phi}_2^2\right)\left<\phi^2\right>+
\tilde{\rho}^2\left<(\nabla^2\phi)^2\right>
\right.\right.\nonumber\\
&&\left.\mbox{}+2\tilde{\phi}_1\tilde{\phi}_2\left<\phi(\bi{x})\phi(\bi{y})\right>
+2\tilde{\rho}\tilde{\phi}_1\left<\phi(\bi{x})\nabla^2\phi(\bi{z})\right>
+2\tilde{\rho}\tilde{\phi}_2\left<\phi(\bi{y})\nabla^2\phi(\bi{z})\right>\right.\nonumber\\
&&\left.\mbox{}-\frac{2}{\left<(\nabla\phi)^2\right>}\left(\tilde{\phi}_1\left<\phi(\bi{x})\nabla\phi(\bi{z})\right>
+\tilde{\phi}_2\left<\phi(\bi{y})\nabla\phi(\bi{z})\right>\right)^2
\right)\nonumber\\
&&\left.\mbox{}+\rmi\left(\tilde{\phi}_1\phi_1+\tilde{\phi}_2\phi_2+\tilde{\rho}\rho\right)\right)
\phi_1\phi_2\rho^2.
\end{eqnarray}
The variables $\phi_1, \phi_2, \rho$ are still Gaussian variables, with,
however, modified correlations which can be found
from (\ref{reduc}).
Therefore, using Wick's theorem
\begin{eqnarray}
\lefteqn{(*)=\frac{2}{\pi\left<(\nabla\phi)^2\right>}\left<\phi(\bi{x})\nabla^2\phi(\bi{z})\right>
\left<\phi(\bi{y})\nabla^2\phi(\bi{z})\right>}\nonumber\\
&&\mbox{}+\frac{\left<(\nabla^2\phi)^2\right>}{\pi\left<(\nabla\phi)^2\right>}
\left(\left<\phi(\bi{x})\phi(\bi{y})\right>
-\frac{2\left<\phi(\bi{x})\nabla\phi(\bi{z})\right>\cdot
\left<\phi(\bi{y})\nabla\phi(\bi{z})\right>}{\left<(\nabla\phi)^2\right>}\right).
\end{eqnarray}
One of these terms is cancelled by (see chapter \ref{vari})
\begin{equation}
\left<\phi(\bi{x})\phi(\bi{y})\right>\left<(\nabla^2\phi)^2\delta(\nabla\phi)\right>
=\left<\phi(\bi{x})\phi(\bi{y})\right>
\frac{\left<(\nabla^2\phi)^2\right>}{\pi\left<(\nabla\phi)^2\right>}
\end{equation}
yielding the following contribution of the measure term:
\begin{equation}
G_1(p)=-\frac{u}{2\pi\left<(\nabla\phi)^2\right>}\left(
\frac{1}{(p^2+\tau)^2}
-\frac{\left<(\nabla^2\phi)^2\right>}{\left<(\nabla\phi)^2\right>p^2(p^2+\tau)^2}\right).
\end{equation}
The vertex function reads
\begin{eqnarray}
\Gamma_2(p)&=&(G_0+G_c+G_1)^{-1}=p^2(p^2+\tau)+u\,\Delta\tau\,p^2\nonumber\\
&&\mbox{}+\frac{u}{2\pi\left<(\nabla\phi)^2\right>}
\left(p^4-\frac{\left<(\nabla^2\phi)^2\right>}{\left<(\nabla\phi)^2\right>}p^2\right)+\Or (u^2)
\nonumber\\
&\equiv&A_{\mbox{\tiny eff}}\left(p^4+\tau_{\mbox{\tiny eff}}\,p^2\right)
\end{eqnarray}
with the effective coupling constants (up to order $\Or (u^2)$ and cutoff momentum $\Lambda=1$)
\begin{equation}
A_{\mbox{\tiny eff}}=1+u/\left(\frac{1}{2\pi}\int_{|p|<\Lambda} \rmd^2p\,\frac{1}{p^2+\tau}\right)=1+\frac{2u}
{\log(1+1/\tau)}
\end{equation}
and
\begin{eqnarray}
\tau_{\mbox{\tiny eff}}&=&\tau+u\,\Delta\tau-\frac{\tau u}{2\pi\left<(\nabla\phi)^2\right>}-
\frac{u\left<(\nabla^2\phi)^2\right>}{2\pi\left<(\nabla\phi)^2\right>^2}\nonumber\\
&=&\tau+u\,\Delta\tau-\frac{2u}{\left(\log(1+1/\tau)\right)^2}.
\end{eqnarray}

\end{document}